\documentclass{sig-alternate-05-2015}
\usepackage{graphicx}
\usepackage{url}
\graphicspath{{img/}}

\usepackage{tabularx, booktabs}
\newcolumntype{Y}{>{\centering\arraybackslash}X}

\usepackage{breakcites}
\usepackage{algorithm2e}

\usepackage{amsmath}
\usepackage{commath}

\usepackage[justification=centering]{caption}
\usepackage{listings}

\makeatletter
\def\@copyrightspace{\relax}
\makeatother

\begin{document}






%

\title{Outlier detection on network flow analysis}
%
%
%
%
%

\numberofauthors{1} 
%
\author{
%
%
\alignauthor
Quang-Vinh Dang\\
       \affaddr{TMC Data Science}\\
       \affaddr{High Tech Campus 96}\\
       \affaddr{NL-5656 AG Eindhoven, the Netherlands}\\
       \email{vinh.dang@tmc.nl}
}

\maketitle
\begin{abstract}
It is important to be able to detect and classify malicious network traffic flows such as DDoS attacks from benign flows. Normally the task is performed by using supervised classification algorithms. In this paper we analyze the usage of outlier detection algorithms for the network traffic classification problem.
\end{abstract}

%
%

\ccsdesc[500]{Computer systems organization~Embedded systems}
\ccsdesc[300]{Computer systems organization~Redundancy}
\ccsdesc{Computer systems organization~Robotics}
\ccsdesc[100]{Networks~Network reliability}

%
%

%
%


\keywords{outlier detection; DDoS attack}

\section{Introduction}

A denial-of-service (DoS) attack  is characterized ``by an explicit
attempt by attackers to prevent the legitimate use of a service" \cite{DBLP:journals/ccr/MirkovicR04}. If the attackers coordinate the DDos traffic from multiple sources to perform the attack, it will be Distributed Denial-of-Service (DDos) \cite{DBLP:conf/icoin/KoayCWS18}.

Multiple studies have analyzed the detection and prevention strategies of DDoS attacks by using classification algorithms \cite{DBLP:journals/cn/DouligerisM04, DBLP:conf/tsp/FouladiKA16, DBLP:conf/icoin/KoayCWS18, DBLP:conf/ntms/AlsirhaniSB18}. While these methods achieved a lot of success, they suffer from imbalanced dataset problem \cite{DBLP:journals/pai/Krawczyk16} and lack of detecting unfamiliar flows. For instance, these techniques mail fail to detect a new DDoS attack technique that they did not see during the training period. Furthermore, supervised classification algorithms usually exhaust of data.

Outlier detection algorithms \cite{DBLP:books/sp/Aggarwal2017} try to distinguish outlier points from normal traffic data. Hence, the techniques might be performed in unsupervised manner \cite{DBLP:journals/datamine/CamposZSCMSAH16}. Furthermore, outlier detection algorithms can deal well with extremed imbalanced dataset, such as 1:1000 ratio \cite{DBLP:journals/pai/Krawczyk16}.

In this paper we evaluate the performance of outlier detection algorithms on detecting DDos traffic. Our work is related to other evaluation studies, such as of \cite{DBLP:journals/corr/abs-1805-00811}. However, the authors of \cite{DBLP:journals/corr/abs-1805-00811} do not analyze the performance in case of imbalanced datasets.

\section{Evaluation}

\subsection{Algorithms}

We evaluated the following algorithms:

\begin{itemize}
    \item \textbf{CBLOF} (Clustering-Based Local Outlier Factor) \cite{DBLP:journals/prl/HeXD03}.
    \item \textbf{HBOS} (Histogram-Based Outlier Score) \cite{goldstein2012histogram}.
    \item \textbf{IForest} {Isolation Forest} \cite{DBLP:conf/icdm/LiuTZ08}.
    \item \textbf{k-NN} (k Nearest Neighbors)\cite{DBLP:conf/sigmod/RamaswamyRS00}.
    \item \textbf{MCD} (Minimum Covariance Determinant) \cite{DBLP:journals/technometrics/RousseeuwD99}.
    \item \textbf{OCSVM} (One-Class SVM) \cite{ma2003time}.
    \item \textbf{PCA} (Principal Component Analysis) \cite{shyu2003novel}.
\end{itemize}

\subsection{Dataset}

We used the dataset provided by \cite{ali2018data} that contains $464,976$ samples that are assigned the labels ``Attack" or ``Benign". We consider these labels as ground truth. A set of 76 features is provided. Some of the features are:

\begin{itemize}
    \item  Flow.Duration
    \item Tot.Fwd.Pkts (Total Forward Packets)
    \item Tot.Bwd.Pkts (Total Backward Packets)
    \item Fwd.Pkt.Len.Max
    \item Fwd.Pkt.Len.Min
\end{itemize}

The ``Attack" traffic contains $3.76\%$ of the whole dataset ($17,462$ samples). In order to evaluate the performance of the outlier detection algorithms, we create different datasets with different ``Attack" Ratio, range from $0.01$ to $0.99$.

To change the class ratio, we applied the following function to select a subset of the dataset:

\begin{lstlisting}[language=python]
def make_dataset(benign_ratio = 0.9, 
        df_attack = df_attack, 
        df_benign = df_benign):
    N_attack = len (df_attack.index)
    N_benign = len (df_benign.index)
    
    n_attack = N_benign / 
        (benign_ratio / (1-benign_ratio))
    sample_df_attack = df_attack.
        sample(int(n_attack),
        replace=False)
    res = pd.concat([df_benign, sample_df_attack])
    res = shuffle (res)

    res = res[[col for col in res 
        if not len(set(df[col]))==1]]

    #divide train/test
    train = res.sample(frac = 0.7, random_state=200)
    test = res.drop(train.index)
    return (train,test)
\end{lstlisting}

\section{Results}

In this section we present the performance in term of $AUC$ and $Accuracy$ scores of each algorithms. Overall, while the $benign\_ratio$ increases the $Accuracy$ scores increase for all algorithms, that can be explained by the imbalanced of the dataset. Another observation is that the performance of the algorithms are very similar between training and testing set due to the fact that these algorithms are all unsupervised. In term of $AUC$ score we could see that IForest and PCA algorithms achieved the best scores while the $benign\_ratio$ increases.

\begin{figure}[h!]
    \centering
    \includegraphics[width = \linewidth]{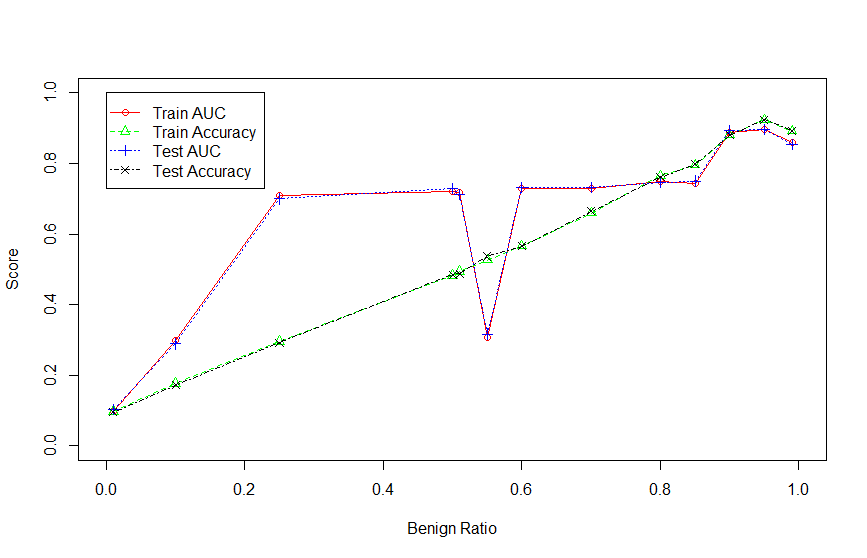}
    \caption{CBLOF}
    \label{fig:cblof}
\end{figure}

\begin{figure}[h!]
    \centering
    \includegraphics[width = \linewidth]{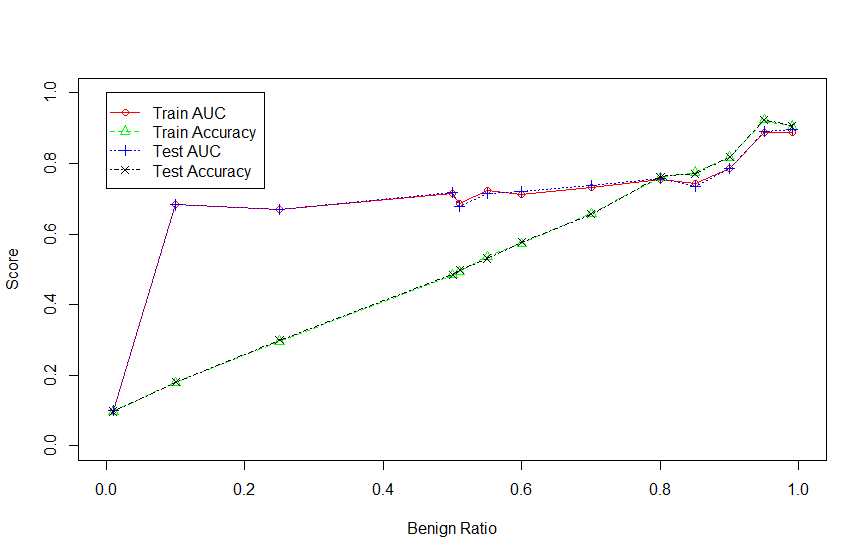}
    \caption{HBOS}
    \label{fig:hbos}
\end{figure}

\begin{figure}[h!]
    \centering
    \includegraphics[width = \linewidth]{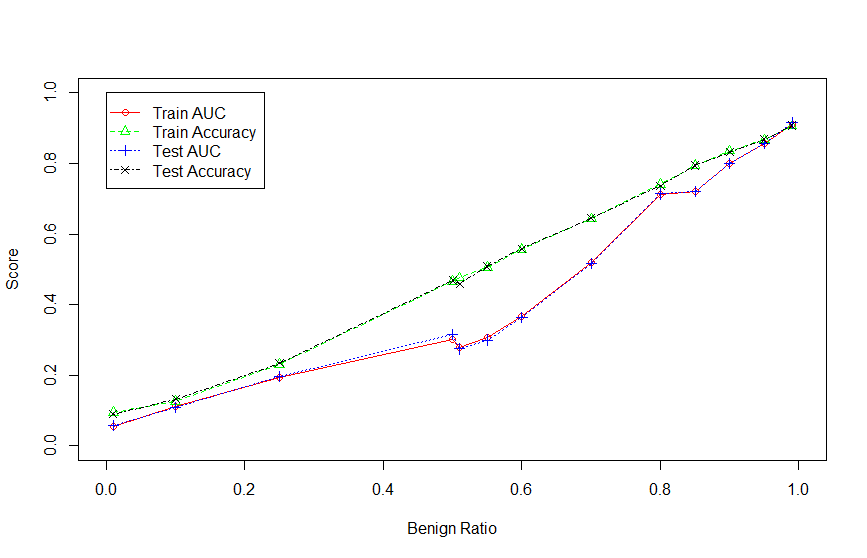}
    \caption{IForest}
    \label{fig:iforest}
\end{figure}

\begin{figure}[h!]
    \centering
    \includegraphics[width = \linewidth]{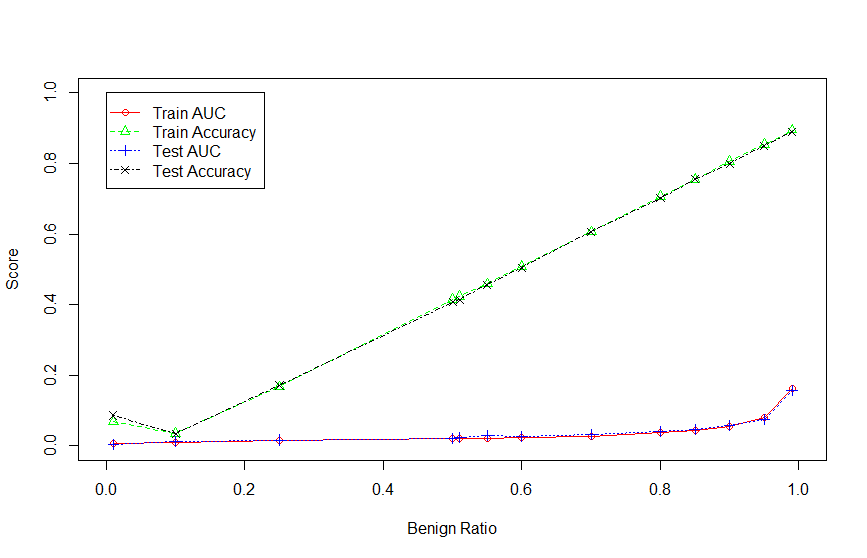}
    \caption{kNN}
    \label{fig:knn}
\end{figure}

\begin{figure}[h!]
    \centering
    \includegraphics[width = \linewidth]{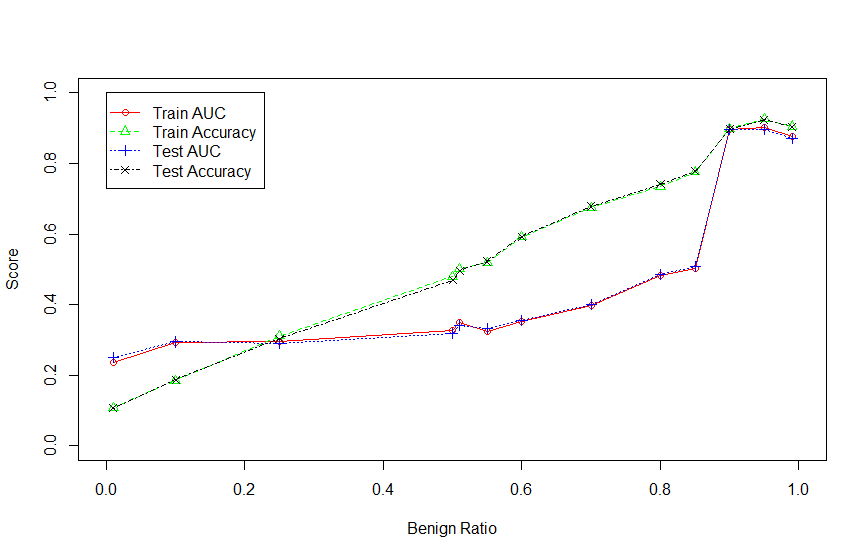}
    \caption{MCD}
    \label{fig:mcd}
\end{figure}

\begin{figure}[h!]
    \centering
    \includegraphics[width = \linewidth]{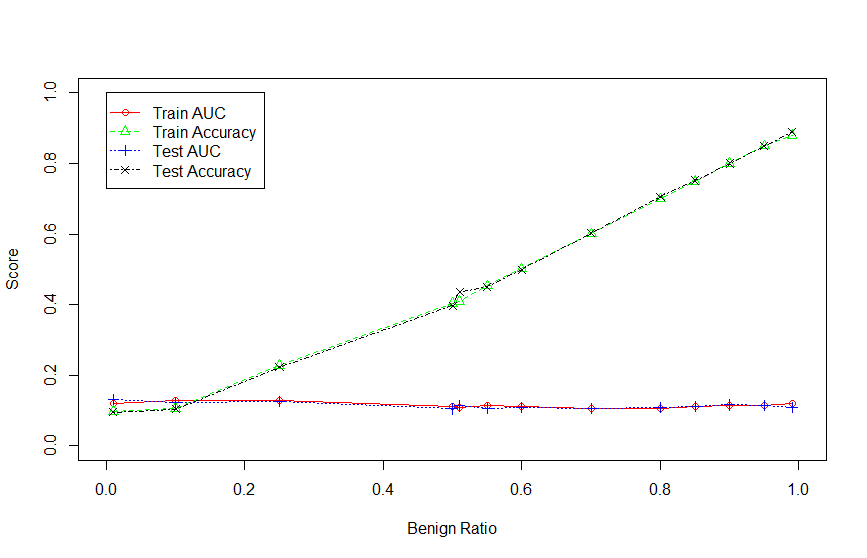}
    \caption{OCSVM}
    \label{fig:ocsvm}
\end{figure}

\begin{figure}[h!]
    \centering
    \includegraphics[width = \linewidth]{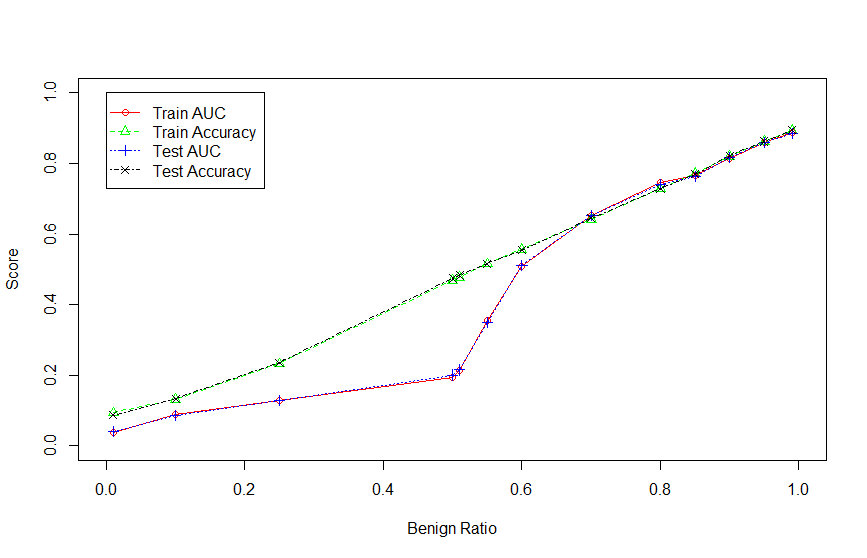}
    \caption{PCA}
    \label{fig:pca}
\end{figure}

\section{Conclusions}

In this paper we evaluate the performance of unsupervised outlier detection algorithms in detecting DDoS attacks. We showed that the outlier detection algorithms perform, particularly Isolation Forest or PCA-based algorithms, perform best if the proportion of outlier instances is small. This contrasts with popular classification algorithms. In the future we will focus on analyzing Isolation Forest and PCA algorithms in other scenarios.

%
\bibliographystyle{abbrv}
\bibliography{sigproc}  
\end{document}